\documentstyle[aps,twocolumn]{revtex}
\input epsf
\def\VOID#1{}

\newcommand{\crg}{CeRu$_{2}$Ge$_2$~}

\def\ceruges{CeRu$_2$Ge$_2${ }}

\begin{document}
%%\begin{frontmatter}
\draft \preprint{V--1.3}
\title{Calorimetric Investigation of CeRu$_2$Ge$_2$ up to 8~GPa}

\author{F. Bouquet, Y. Wang, H. Wilhelm, D. Jaccard and A. Junod}
\address{D\'epartement de Physique de la Mati\`ere Condens\'ee,
  Universit\'e de Gen\`eve,\\ 24, Quai E.--Ansermet, CH--1211 Geneva
  4, Switzerland}
\maketitle \widetext

\begin{abstract}
  We have developed a calorimeter able to give a qualitative picture
  of the specific heat of a sample under high pressure up to $\approx
  10$~GPa. The principle of {\sc ac}--calorimetry was adapted to the
  conditions in a high pressure clamp. The performance of this
  technique was successfully tested with the measurement of the
  specific heat of \ceruges in the temperature range $1.5$~K~$ < T <
  12$~K. The phase diagram of its magnetic phases is consistent with
  previous transport measurements.
\end{abstract}
%%\pacs{ Keywords: A.~magnetically ordered material, D.~phase transition, D.~heat capacity, E.~high pressure.}
%%\end{frontmatter}

\narrowtext
\tighten
\section{Introduction}
Ambient pressure calorimetry is a classical tool for condensed matter
studies. The knowledge of specific heat and its temperature dependence provides
information on the type of solid--state excitations and on phase transitions.
Extending the standard methods to high pressures (up to $\approx 10$~GPa) is
a rather difficult and experimentally challenging task. In a high pressure
study, tiny samples are embedded in a pressure transmitting medium. This
assembly is enclosed in a pressure cell. The unavoidable thermal contact
between sample and environment leads to heat leakage problems in
conventional calorimetric techniques. If the heat capacity of the empty high
pressure device is known, the heat capacity of the sample under pressure can
be determined by subtraction~\cite{Ho66,Phill68}. As this will work in
practice only if the sample contributes a considerable part to the total heat
capacity, the choice of samples is restricted.

In the present work we report how the {\sc ac}--technique~\cite{Sullivan} can
be used to determine the specific heat at low temperature and high pressure.
Indeed, this technique is known to be well suited to the high pressure
environment \cite{Eichler,Garland} since it does not require adiabatic
conditions, and allows high resolution, even on very small samples. However,
absolute accuracy is generally difficult to obtain.

Two different sample arrangements within the clamped high pressure device
were tested. CeRu$_2$Ge$_2$ was chosen as the sample. At ambient pressure it
shows two magnetic phase transitions \cite{WILHE98,RAYMO99} which give rise
to large signatures in the specific heat ($C/T\approx 8$~and 0.5~J/molK$^2$
at the Curie $T_{\rm C}$ and N\a'eel temperature $T_{\rm N}$, respectively).
The pressure dependence of the transition temperatures is known from
transport measurements \cite{WILHE98,WILHE99,KOBAY98,SUELL99,WILHE99a}. Thus,
\crg is a good candidate for testing the {\sc ac}--calorimetric technique at
high pressure. It is expected that specific heat can provide additional
information on the magnetic phase transitions.
\section{AC--Calorimetry}
\begin{figure}
\epsfxsize=150mm \epsfbox{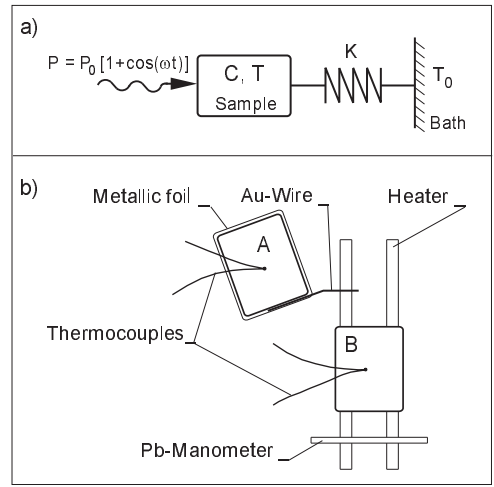}
  \vspace*{-135mm}
  \caption{
    a) Schematic description of {\sc ac}--calorimetry.  b) Arrangement
    of the samples in the high pressure cell. Sample A is placed on
    top of the heater wires but is insulated from them. Sample B is
    in contact with a metallic foil (Pb). The
    Chromel--\underline{Au}Fe thermocouples measure the sample
    temperature. The Pb--wire serves as pressure gauge $[$13$]$.}
  \label{fig:schema}
\end{figure}

The  principle of {\sc ac}--calorimetry is described in Ref.~\cite{Sullivan}.
Figure~\ref{fig:schema}(a) shows a simplified model: the sample is ther-
\newpage
\vspace*{22mm}
\noindent
mally excited by a {\sc ac}--heater, and the amplitude of the temperature
oscillations $T_{ac}$ is measured. In the ideal case, when the heat capacity
of thermometer, heater, and heat link are negligible and when the coupling
between heater, sample, and thermometer is ideal, $T_{ac}$ depends
\noindent
only on the working frequency $\omega$,  the specific heat of the sample $C$, and the
global heat link $K$, through the equation:
\begin{equation}
T_{ac} = \left| \frac{P_0}{K + iC\omega}  \right| ,
\label{eq:tac}
\end{equation}
with $i^2 =-1 $ and $P_0$ the mean heater power.  When working
above the cut-off frequency $\omega_1 = K/C $, $T_{ac}$ is inversely
proportional to $C\omega$.  The possibility of tuning both the
amplitude and the frequency of the excitation is the main advantage of
this technique; as long as $K$ can be made small enough, the
sensitivity of the measurement does not depend on the mass of the
sample.

The real case of a sample in a pressure cell is far from this ideal
one. In particular the specific heat of the pressure transmitting
medium has to be taken into account as was done by Baloga and
Garland~\cite{Garland}. However, the general behaviour of
Eq.(\ref{eq:tac}) can be recovered if the product $c \lambda$ (volumic
specific heat times heat conductivity) of the pressure transmitting
medium is negligible with respect to that of the
sample~\cite{Garland}. If so, the heat wave does not propagate too
far into the medium ($\lambda$ small) and the specific heat of the
temperature oscillating medium does not contribute too much ($c$
small). Hence, thermal properties of the pressure transmitting medium
determine the working conditions.

\section{Experimental Details}

Calorimetric measurements were performed in a clamped high pressure device
able to reach $\approx$~10~GPa.  Sample preparation and details of the high
pressure cell are given in Ref.~\cite{JACCA98,WILHE99}. An important point
for specific heat measurements is the fact that the samples are embedded in
steatite, the pressure transmitting medium. Figure~\ref{fig:schema}(b) shows
schematically the inner part of the high pressure cell in detail. The typical
thicknesses of the sample, thermocouple and heating wire are 20, 12, and
3~$\mu$m, respectively. Two different ways of supplying the heat to the
samples were tested. For sample~A a thin electrical insulation (4--5~$\mu$m
of an epoxy/Al$_2$O$_3$ mixture) prevents electrical contact with the heater
but still allows a good thermal contact. Sample B is set apart on a metallic
(Pb) foil, electrically (and thus thermally) linked to the heater through a
gold wire.  No heating current passes through this sample.

Temperature oscillations should be small compared to the sample
temperature and were chosen in the range 2~mK~$< T_{ac} <$~20~mK. This
gives thermovoltages of $\approx$~100~nV, which were amplified at room
temperature in two stages ($\times$~500) and finally read by a
lock--in amplifier. We used frequencies between 500 and 4000~Hz.  The
noise after amplification was 0.15~$\mu$V peak--peak, i.e.
0.75~nV$/\sqrt{\rm{Hz}}$~rms, referred to the thermocouple input
taking into account the amplification and settings of the lock--in
amplifier. This is not far from the Johnson noise
(0.3~nV$/\sqrt{\rm{Hz}}$~rms) of the connecting wires (5~$\Omega$ at
300~K).  The typical thermopower $S$ of the thermocouple is
10~$\mu$V/K, which gives a temperature noise for the sample of
30~$\mu$K peak--peak; to reach a sensitivity of 1\% in the
determination of $C$, the sample should have a temperature oscillation
$T_{ac} = $~3~mK.

The resistivity of the Pb manometer was measured by the four point
technique at 72~Hz. The current excitation was low (5~$\mu$A) to avoid
local heating.

\section{Results and Discussion}
\subsection{Testing}
\begin{figure}
\epsfxsize=90mm \epsfbox{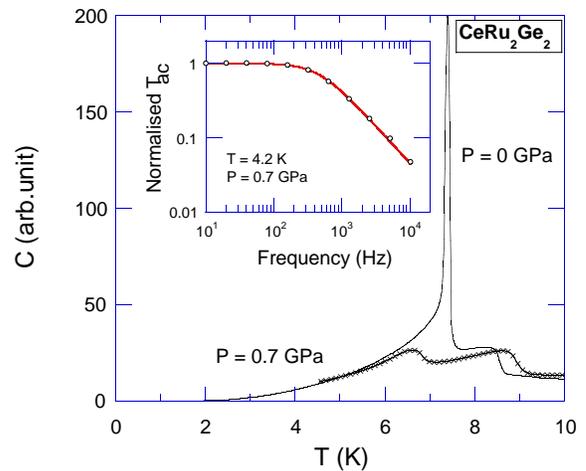}
  \vspace*{-55mm}
  \caption{
    Low pressure specific heat measurement compared with that at ambient
    pressure. The curves are normalized at 10~K.  The inset shows a frequency
    test done at 4.2~K and 0.7~GPa.  The solid line is a low pass
    filter fit with a cut-off frequency of 450~Hz.}
  \label{fig:1eremesure}
\end{figure}

In order to confirm that the signal truly reflects the heat capacity, it
should obey the frequency dependence described by the general formula given
in Ref.~\cite{Garland}.  However, the simplified Eq.(\ref{eq:tac}) describes
well the signal as can be seen in the inset of Fig.~\ref{fig:1eremesure}. At
low temperature (4.2~K) and pressure (0.7~GPa) the cut-off frequency is
$\approx$~450~Hz; the working frequency was chosen to be slightly
higher, keeping in mind that the signal decreases with $\omega$.  Changing
the temperature and the pressure influences the cut-off frequency: from 1 to
4~K we worked in the range 600--1000~Hz, and from 4 to 12~K between 2000 and
4000~Hz.

The specific heat of the sample is certainly the main contribution to the
signal. This follows implicitly from the fact that the frequency dependence
is well described by Eq.(\ref{eq:tac}) without any correction. However, a
convincing argument is presented in Fig.~\ref{fig:1eremesure}, where the
specific heat measured at ambient and low pressure (0.7~GPa) are
compared. The former curve was obtained with a conventional relaxation
technique using a comparatively large sample (14~mg)~\cite{WILHE99,Bernard}
whereas the investigated samples each had a mass of $\approx$~10~$\mu$g. The
phase transitions are clearly visible.  They are also detectable with the
{\sc ac}--technique at 0.7~GPa, but slightly shifted in temperature as
expected from the phase diagram
\cite{WILHE98,WILHE99,KOBAY98,SUELL99,WILHE99a}.  The height of the specific
heat jump at the second order transition ($T_{\rm N} \approx$~9~K) represents
47\% of the total signal compared to 51\% for the ambient pressure curve.
This indicates that the heat capacity measured under pressure is in majority
the heat capacity of the sample.

The first order transition ($T_{\rm C} \approx$~7~K) is not a good
reference for such a comparison. The height of its peak is very
sensitive to any distribution of $T_{\rm C}$. Moreover, {\sc
  ac}--calorimetry is not the proper tool to measure a latent
heat~\cite{aclatentheat1,aclatentheat2,aclatentheat3,Bouquet}, since
it only detects the reversible part at  frequency $\omega$ on a
temperature scale  $T_{ac}$.  The apparent latent heat may be
smaller if part of the system is irreversible at the given frequency.
The size of the peak can depend strongly on the operating
conditions~\cite{Bouquet}.  This explains the different behaviour at
$T_{\rm C}$ since the relaxation technique used at zero
pressure~\cite{WILHE99,Bernard} is not subject to this restriction.
However, the position of a first order transition can be detected by an
{\sc ac}--calorimeter.

This setup allows the specific heat of samples under pressure to be measure
almost without any addenda, therefore no background has to be subtracted. But
there are two main limitations: (i) the temperature of the samples is
measured with thermocouples, with the assumption that their calibration does
not change under pressure.  Hence, the quantitative comparison of data at
different pressures is limited.  However, the relative uncertainty of data
obtained in a narrow  pressure range remains small. (ii) The total amount of
power supplied to the samples is not known precisely, although a joule heater
was used.  The measurements were done under the assumption that the heat
power received by the samples does not change with pressure. These
limitations presently do not allow the acquisition of absolute values for the
specific heat, so we have to rely on a separate measurement done at zero
pressure.

\subsection{Specific Heat of \crg at high pressure}

\begin{figure}
\epsfxsize=90mm \epsfbox{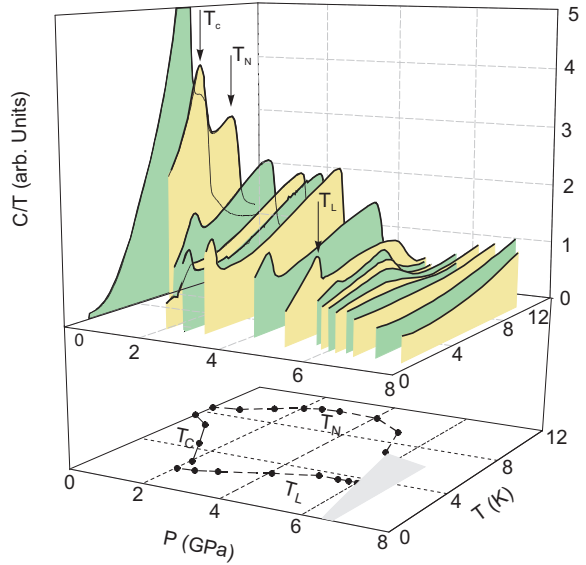}
  \vspace*{-35mm}
  \caption{
    $C/T$ vs. $T$ and ($T$,$P$) phase diagram of CeRu$_2$Ge$_2$. The
    $C/T$ curves below (above) 5~GPa represent sample A(B). Below
    2~GPa the AFM and FM ordering give pronounced signatures in the
    specific heat. A new phase transition (at $T_{\rm L}$) occurs
    around 2~K at $P=2.0$~GPa.  Magnetic order is suppressed above
    $P_c=6.9\pm 0.3$~GPa. The lower part shows the deduced pressure
    dependence of the ordering temperatures. In the dashed region, the
    uncertainties in the determination of $T_{\rm N}$ are large (see
    text). }
  \label{fig:3D}
\end{figure}

Figure~\ref{fig:3D} shows $C/T$ versus $T$ curves of \crg at different
pressures and the derived ($T$,$P$) phase diagram. The evolution of the anomalies in
$C/T$ can be followed up to high pressure ($P<8$~GPa) for the first time.
Results for sample A (insulated from the heater and on top of it) and sample
B (connected to the heater but set farther away) at low ($\leq 5$~GPa) and
high pressure, respectively, were used for this figure.  Indeed, both samples
show almost identical results, but present small differences which explain
this choice. The anomalies in sample B tend to be $\approx$~10\% smaller
(depending on pressure and temperature) since the heat capacity of the
metallic foil contributes to the measured signal, too. On the other hand, the
anomalies of the specific heat of sample A tend to be broader, probably due
to a deviation from hydrostatic pressure conditions, especially at higher
pressure.

The phase diagram obtained compares well with that deduced from transport
measurements \cite{WILHE98,WILHE99,KOBAY98,SUELL99}. The N\a'eel temperature
first increases with pressure up to $T_{\rm N}=11$~K (at $P=3$~GPa) and then
decreases. In contrast to this, the Curie temperature decreases with pressure
and vanishes around 2~GPa.  Above that pressure \crg enters a differently
ordered magnetic groundstate below a characteristic temperature $T_{\rm
L}\approx 2$~K. This transition temperature increases up to $\approx 3.5$~K
(at $P=5.0$~GPa). Magnetic order seems to be suppressed below 1.5~K in the
pressure range $6.6<P<7.2$~GPa. In this interval the anomaly related to
$T_{\rm N}$ becomes weaker and the overall shape of the $C/T$ vs. $T$ curves
changes. Hence, a critical pressure $P_c=6.9\pm 0.3$~GPa is inferred. This
value is in agreement with the one deduced from transport measurements
(electrical resistivity \cite{SUELL99}, thermoelectrical power
\cite{WILHE99a}) performed at temperatures above 1.5~K.  Electrical
resistivity measurements carried out down to 30~mK and 60~mK yield
$P_c=8.7$~GPa~\cite{WILHE99} and 10~GPa \cite{KOBAY98}, respectively. These
differences are probably due to the influence of sample quality --- on a
microscopic scale --- even though the samples investigated here are taken from
the same ingot as those used in the work of Ref.~\cite{WILHE99,WILHE99a}.

The $T_{\rm N}(P)$ dependence can be compared qualitatively with the
Doniach phase diagram \cite{DONIA77} which describes the competition
between the Kondo effect and the RKKY interaction. At low pressures,
the latter overcomes the Kondo effect.  As the non-magnetic phase
boundary is reached, the magnetic moment of the localized
$4f$--electrons is screened completely by the spin of the conduction
electrons.

\begin{figure}
\epsfxsize=90mm \epsfbox{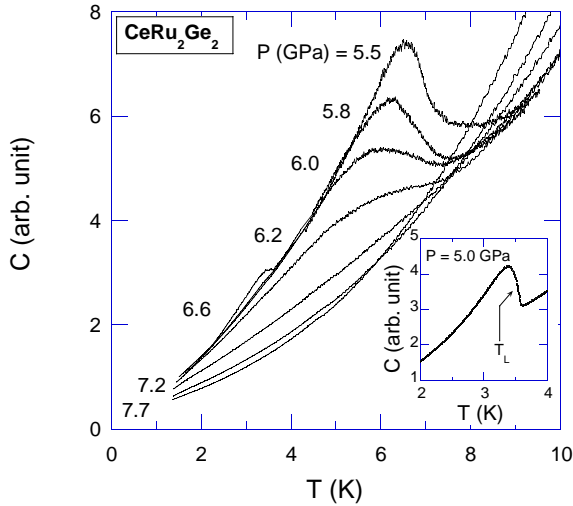}
  \vspace*{-50mm}
  \caption{
    Specific heat $C(T)$ of \crg (sample B) at high pressure.  The
    pronounced feature is related to the antiferromagnetic transition
    (at $T_{\rm N}$) which is suppressed by pressure. The inset shows
    the specific heat at 5.0~GPa where the anomaly at $T_{\rm L}$ is
    maximum. At 5.5~GPa this anomaly can still be seen near
    3.5~K (main frame).  }
  \label{fig:physique}
\end{figure}

A detailed view of the $C(T)$--data recorded above 5.5~GPa is given in
Fig.~\ref{fig:physique}. The broadening of the antiferromagnetic
transition is evident and may be related to an intrinsic behaviour of
\crg or more likely to a deviation from hydrostatic conditions, since
the sample sees a pressure inhomogeneity of about $\pm 0.4$~GPa around
7~GPa.

The inset of Fig.~\ref{fig:physique} shows the low temperature part of
the specific heat measured at 5.0~GPa where the anomaly at $T_{\rm L}$
is maximum.  The specific heat data confirm that this transition, also
observed in transport measurements \cite{WILHE99,WILHE99a}, has
thermodynamic origin and is a bulk property. The nature of this
transition is still unclear. Several neutron experiments on
substituted compounds (substitution of Ge and Ce by Si and La,
respectively, simulates a pressure effect; a correspondence with
pressure is given in Ref.~\cite{WILHE99}) point to the existence of a
low temperature transition even if the order parameter has not been
identified yet. The indications are (i) the increase of the third
order harmonic of the magnetic modulation in CeRu$_2$SiGe, leading to
a squaring of the magnetic modulation \cite{RAINF98}; (ii) a change in
the magnetic excitation spectrum of CeRu$_2$SiGe from quasielastic
above $T_{\rm L}\approx 2$~K to inelastic below $T_{\rm L}$
\cite{RAINF98}; (iii) in Ce$_{0.8}$La$_{0.2}$Ru$_2$Si$_2$ the diffuse
scattering measured slightly off the magnetic Bragg peak shows a
maximum at $T_{\rm L}\approx 1.8$~K instead at $T_{\rm N}\approx
5.6$~K \cite{REGNA90}.

\section{Conclusion}

The specific heat of \ceruges in the temperature range 1.5--11~K was
measured up to 8~GPa with an {\sc ac}--calorimeter. The evolution of
the various magnetic phase transitions in this intermetallic compound
could be followed up to the magnetic to non--magnetic phase boundary
($P_c = 6.9 \pm 0.3$~GPa). The ($T$,$P$) phase diagram is in excellent
agreement with the previously presented one. This demonstrates that
{\sc ac}--calorimetry can be successfully adapted to high pressure
experiments in a clamp pressure device, and opens a new route for
thermodynamic measurements.

%%\begin{ack}
\section*{Acknowledgements}
We thank S. Raymond, P. Haen and R. Calemczuk for helpful discussions.
This work was partly supported by the Swiss National Science
Foundation. During writing the manuscript we were informed of the progress of the
Grenoble group~\cite{Grenoble99} who also successfully adapted {\sc
  ac}--calorimetry to a high pressure device, using a diamond anvil
cell and helium as pressure transmitting medium.
%%\end{ack}

\end{document}